# Advancements in Solid-State Sodium-Based Batteries: A Comprehensive Review

Arianna Massaro,[1,2,*] Lorenzo Squillantini,[3] Francesca De Giorgio,[2,3] Francesca A. Scaramuzzo,[2,4] Mauro Pasquali,[2,4] Sergio Brutti[2,5,*]

[1] *Dipartimento di Scienze Chimiche, Università degli Studi di Napoli Federico II, Compl. Univ. Monte Sant'Angelo – Via Cintia 21, 80126 Napoli, Italy - arianna.massaro@unina.it (ORCID: 0000-0003-2950-6745)*

[2] *National Reference Center for Electrochemical Energy Storage (GISEL), INSTM – Via G. Giusti 9, 50121 Firenze, Italy*

[3] *Consiglio Nazionale delle Ricerche, Istituto per lo Studio dei Materiali Nanostrutturati (CNR-ISMN) – Via P. Gobetti 101, 40129 Bologna, Italy - francesca.degiorgio@cnr.it (ORCID: 0000-0003-3780-6238), lorenzosquillantini@cnr.it (ORCID: 0000-0001-7678-6909)*

[4] *Dipartimento di Scienze di Base e Applicate per l'Ingegneria (SBAI), Sapienza Università di Roma – Via del Castro Laurenziano 7, 00161 Roma, Italy - francesca.scaramuzzo@uniroma1.it (ORCID: 0000-0001-9921-5252), mauro.pasquali@uniroma1.it (ORCID: 0000-0003-4627-4253)*

[5] *Dipartimento di Chimica, Sapienza Università di Roma – P.le Aldo Moro 5, 00185 Roma, Italy - sergio.brutti@uniroma1.it (ORCID: 0000-0001-8853-9710)*

**Synopsis**

**Abstract**

This manuscript explores recent advancements in solid-state sodium-based battery technology, particularly focusing on electrochemical performance and the challenges associated with developing efficient solid electrolytes. The replacement of conventional liquid electrolytes with solid-state alternatives offers numerous benefits, including enhanced safety and environmental sustainability, as solid-state systems reduce flammability and harsh chemical handling. The work emphasizes the importance of structure and interface characteristics in solid electrolytes, which play a critical role in ionic conductivity and overall battery performance. Various classes of solid electrolytes, such as sodium-based anti-perovskites and sulphide electrolytes, are examined, highlighting their unique ionic transport mechanisms and mechanical properties that facilitate stable cycling. The manuscript also discusses strategies to enhance interfacial stability between the anode and the solid electrolyte to mitigate performance degradation during battery operation. Furthermore, advancements in electrode formulations and the integration of novel materials are considered pivotal in optimizing the charging and discharging processes, thus improving the energy and power densities of sodium batteries. The outlook on the future of sodium-based solid-state batteries underscores their potential to meet emerging energy storage demands while leveraging the abundant availability of sodium compared to lithium. This comprehensive review aims to provide insights into ongoing research and prospective directions for the commercialization of solid-state sodium-based batteries, positioning them as viable alternatives in the renewable energy landscape.

## 1. Introduction

The quest for efficient energy storage systems has never been more critical than in the present era, which is intimately characterized by burgeoning energy demands and concerns over environmental sustainability. Among various contenders, rechargeable batteries have emerged as indispensable components, powering a myriad of applications ranging from portable electronics to electric vehicles (EVs). The development and commercialization of lithium-ion batteries (LIBs) have been pivotal in advancing technologies such as electric transportation and renewable energy systems. However, the reliance on a few almost irreplaceable metals, particularly lithium, cobalt, and nickel, poses significant challenges for the sustainability and scalability of LIB technologies. One of the primary questions facing LIB technologies is the limited availability of critical raw materials (CRMs): for instance, lithium and cobalt are essential for next-gen positive electrode manufacturing for LIBs [1–3], yet their supply is fraught with geopolitical risks and ethical concerns related to mining practices [4,5], as well as to the environmental impact of brine processing for $Li_2CO_3$ production [6]. Their increasing demand on the international commodity markets, driven by the rapid expansion of the EV sales, exacerbates these supply risks. While LIBs have revolutionized energy storage technologies, their reliance on CRMs presents significant challenges for sustainability and environmental impact, especially in consideration of the inevitable use of transition metals in positive electrodes and solid-state electrolytes [8–10]. Furthermore, currently ~95% of LIBs are landfilled rather than recycled, indicating a significant gap in the recycling infrastructure necessary to reclaim these critical materials [4]. It is a matter of fact that recycling technologies for LIBs are still in their infancy, and existing methods often fail to recover materials in a form suitable for direct reuse in new batteries. The environmental impact of these recycling processes can be significant, often resulting in higher energy consumption and emissions compared to primary production methods [7]. The current recycling infrastructure is inadequate to meet the growing demand for the materials used in LIBs, necessitating urgent advancements in recycling technologies and the exploration of alternative battery chemistries [11–14]. Addressing these challenges is essential for ensuring the long-term viability of effective energy storage technologies and achieving a sustainable energy future. In this respect, the exploration of beyond-lithium energy storage technologies has gained significant momentum, being the availability of reliable energy storage systems based on alternative and competitive chemistries an effective way to minimize systemic risks by improving the market flexibility [15,16].

## 2. Solid-state sodium batteries: from fundamental concepts and formulations to energetics and sustainability features

Sodium-based batteries (SBs) have emerged as a promising alternative to LIBs and have attracted considerable attention. The appeal of sodium as an electroactive material lies in several key factors. Sodium is abundantly available worldwide, making it a cost-effective and sustainable substitute for lithium, which is subject to price fluctuations and geopolitical concerns. Furthermore, it shares similar electrochemical properties with lithium, allowing for its integration into existing battery technologies with minimal modifications. Its low reduction potential enables high energy densities, thereby enhancing the performance of sodium-based systems. Beside the beneficial impact of the use of sodium instead of lithium, SBs also rely on the manganese-redox chemistry at the cathode side rather than cobalt [17,18]. This is another pivotal advantage compared to LIBs, being Mn commodities widely available worldwide [19]. Among the various configurations of SBs, solid-state batteries have emerged as a particularly attractive option. On one hand, the solid-state electrolyte (SE) offers several advantages over liquid electrolyte counterparts, including enhanced safety, wider operating temperature ranges, and improved stability against dendrite formation, thereby mitigating concerns related to short-circuiting and thermal runaway. On the other hand, the solid-state architecture enables the use of metallic sodium anodes, which further contributes to the overall energy density and cycling stability of the battery. Furthermore, the concept of all-solid-state batteries (ASSBs) represents a paradigm shift in battery technology, promising even greater advancements in terms of safety, energy density, and longevity [20–22]. By employing solid-state electrolytes for both the anode and cathode, ASSBs offer unparalleled levels of safety and stability, making them highly desirable for applications requiring robust and reliable energy storage solutions.

This review paper offers an in-depth analysis of the current advancements in solid-state sodium-based battery systems. By exploring the fundamental principles, performance metrics, and technological challenges of these battery formulations, it seeks to provide critical insights into the development and future direction of next-generation energy storage technologies. The performance of ASSBs is closely linked to the properties of SEs. These electrolytes not only replace flammable liquid electrolytes, thereby improving safety, but also enable the use of high-voltage cathodes and sodium metal anodes, which can lead to higher energy densities [23,24]. The use of sodium metal as anode material is particularly advantageous, given its high theoretical capacity of approximately 1166 mAh • $g^{-1}$ and low electrochemical potential of -2.7 V *vs.* SHE (standard hydrogen electrode), which can significantly contribute to the overall energy performance of these batteries [25,26]. However, challenges such as dendrite formation and compatibility with solid-state electrolytes must be addressed to fully realize their potential [27]. On the positive electrode side, the use of SEs would allow the full exploitation of high-voltage active materials like the NASICON-type cathode material $Na_4MnCr(PO_4)_3$ or $Na_3V_2(PO_4)_2F_3$ (NVPF) and similar phases (*e.g.*, $Na_3(VOPO_4)_2F$, $Na_4Ni_3(PO_4)_2(P_2O_7)$ and $Na_{3.5}V_{1.5}Mn_{0.5}(PO_4)_3$), thanks to the minimization of the side reactivity typically observed in sodium-based liquid electrolytes [28–32]. These materials exploit redox processes between 3.7 and 4.2 V *vs.* $Na^+$/Na, along with a sustained reversible capacity of 100-130 mAh • $g^{-1}$ and a specific energy of 500 Wh • $kg^{-1}$.

The major bottleneck that hinders the development of effective ASSBs is represented by inadequate physical-chemical properties at RT of any SE so far proposed in the available literature (generally the $Na^+$ conductivity in conventional liquid electrolytes reaches few mS • $cm^{-1}$ order of magnitude, while solid counterparts usually exhibit values in the range of $10^{-5}$-$10^{-1}$ S • $cm^{-1}$ depending on the chemical nature of the solid electrolyte, *i.e.*, polymer or ceramics). As discussed below in a dedicated section, the low ionic conductivities delivered by SEs at RT limits the performance of ASSBs in respect to the state-of-the-art SBs based on liquid aprotic electrolytes. On the other hand, the intrinsic advantages from the safety and manufacturing points of view of ASSBs in respect to standard SBs formulations is driving the worldwide efforts in public and private R&D laboratories to deliver advancements and breakthroughs in the search for effective SE formulations. Overall, different families of SEs, including ceramics [33], polymers [34], and composites [35,36], can be framed within the ASSBs formulations. Each family has specific advantages and disadvantages from the point of view of transport properties, mechanical stability, manufacturability and interface formation. This last point, *i.e.*, the interface stability and its efficacy in the charge transport, is pivotal for any SE as its local failure or misfunctioning easily leads to performance drop of the entire device [37–39]. Thus, the development and optimization of an effective solid-solid interface is a pre-requisite for the technology validation of any ASSBs. In the literature, a variety of practical strategies have been tested to tackle both the charge storage and transport mechanism across the intrinsically bad contacts generated at the boundary of heterogenous junctions. Inevitably, for each specific cell configuration, targeted solutions have been outlined to improve the quality of metallic sodium crystal growth and dendrites, the crystalline mismatch at the interface, the volume changes upon stripping/deposition, as well as the growth of additional interlayers between the SE and the Na-metal layer [40].

Recently, increasing attention to the validation of the use of SEs in Na-S and Na-$O_2$ batteries have been reported [41], where the solid-state phase turns out to efficiently impact the electrochemical performance by limiting the diffusion of polysulphides and the sodium dendrite formation in the former [42], and by reducing the electrolyte decomposition and thus preventing the formation of insoluble by-products in the latter formulation [43].

While all the abovementioned key enablers towards the ASSBs development will be analytically addressed in the following sections, we would like to stress that sustainability must be at the centre stage in the development of ASSBs to allow the flourishing of a technology rooted into the circular economy and environmental paradigms, as pictorially represented in Fig. 1. To this aim, in this work, we will review the current state-of-art through these special "sustainability" lenses.

As anticipated, the Li-to-Na shift offers a compelling case for an improved sustainable technology compared to competitors. Sodium, being far more abundant and more evenly distributed in the Earth's crust than lithium, alleviates concerns over resource scarcity and geopolitical dependencies, while its lower cost fosters broader adoption in large-scale applications. What is even more enhanced here is that transitioning from liquid to solid-state configurations also carries on the minimization of electrolyte amounts and the accompanying safety hazards. As a matter of fact, SE films are typically constrained to thickness below 100 μm, scarcely flammable and non-spillable. SEs would also mitigate reliance on fluorinated compounds commonly found in liquid electrolytes (*e.g.*, $NaPF_6$ salt is generally present), aligning with the regulatory PFAS directive to phase out any per- and polyfluoroalkyl substance due to their environmental persistence and toxicity. These advantages position all-solid-state sodium batteries as a transformative technology that not only meets performance demands but also aligns with global sustainability goals. Moving forward, innovation in materials science, coupled with thoughtful policy alignment, will be essential to unlocking the full potential of this promising battery technology.

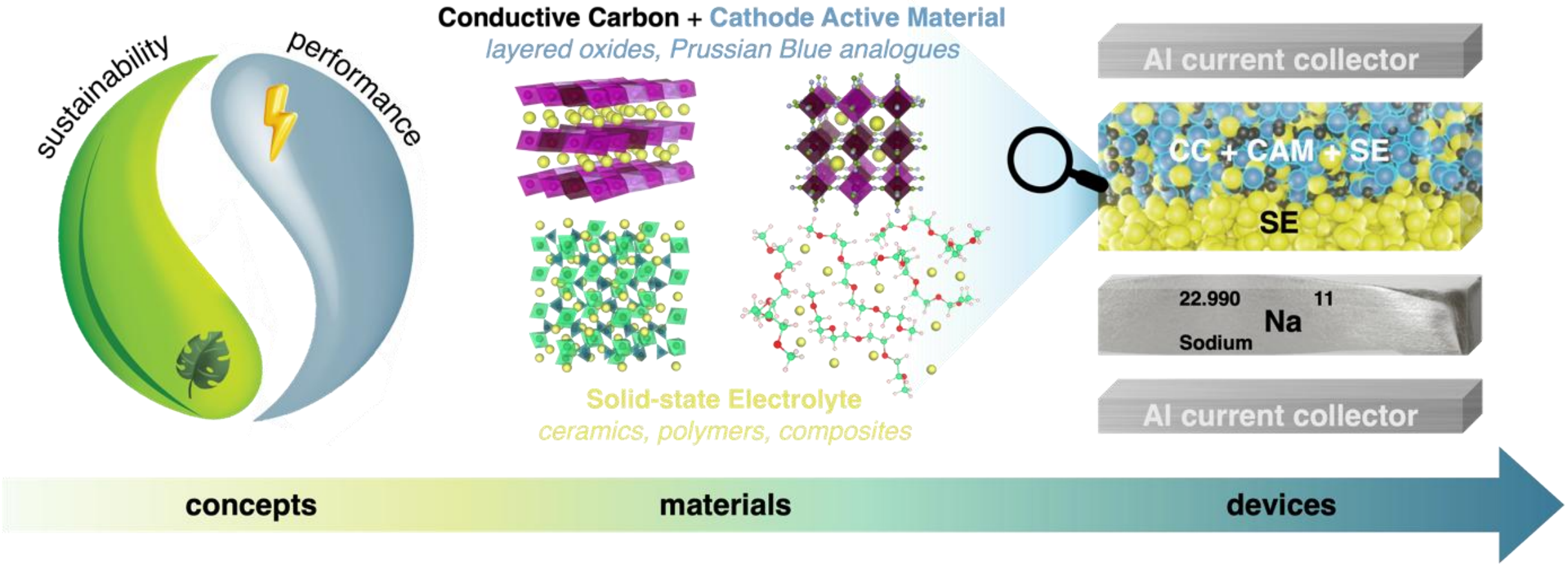

***Figure 1.*** *New paradigm for the development of future sodium-based ASSBs: the balance between the two fundamental concepts, sustainability and performance, needs to be reflected in the design of efficient functional materials, including cathode active materials and SEs, which would ultimately lead to the desired device configurations.*

## 3. Key enabler in the liquid-to-solid transition: challenges and advances for solid-state electrolytes

ASSBs promise improved operational safety as well as enhanced energy and power density that are essential for deployment in large-scale electric grids. The effective replacement of organic liquid electrolytes and porous polymer separators in state-of-the-art battery configurations calls for well-designed solid ion conductors. Overall, the motion of ions between the anode and the cathode (*e.g.*, $Li^+$ or $Na^+$) can be achieved by a variety of SEs that offer relevant environmental advantages owing to their reduced flammability, prevention of solvent leakage, as well as easier recycling and recovery procedures [44–47]. Several requirements should be fulfilled to achieve adequate electrochemical performance with any SE (*e.g.*, large ionic conductivity, large $Na^+$ transference number, small anodic/cathodic charge transfer resistances, large electrochemical stability window, large thermal stability, limited chemical reactivity towards sodium metal/de-sodiated anodes/de-sodiated cathodes), and great research efforts are currently devoted to identifying viable SE candidates and tuning their structure-property relationship towards effective application in sodium cells.

While liquid electrolytes can penetrate the porous structure of the electrodes, the solid electrolyte-electrode interface may exhibit an unoptimized surface contact, leading to inappropriate electrochemical compatibility, sluggish charge transfer kinetics, and small ionic conductivity [48]. Values of at least 1 mS • $cm^{-1}$ near RT are commonly targeted to compete with established organic liquid electrolytes and ensure suitable charging and discharging rates. Indeed, interface stability is particularly relevant and should be pursued to withstand also challenging electrochemical conditions, such as the very high/low voltage range where ASSBs are generally operated. Additional physical-chemical and electrochemical properties, including electron-insulating character and wide electrochemical stability window, are always required to avoid self-discharge, short circuiting, and undesired degradation processes. Still, thermal stability should not be neglected in view of easily manufacturing processes and sustainable operating conditions.

A large variety of materials have been proposed as suitable candidates in practical applications and in different battery designs according to their specific mechanical and electrochemical properties. While general considerations are addressed somewhere else [45], here our aim is to review the most emerging families of Na-ion conductors and focus on their central role in the ongoing and future research for solid-state sodium batteries as next-generation energy storage technology. We would like to stress that the development of innovative materials for real SEs requires to consider also additional descriptors to parallel the most relevant functional properties of any battery electrolyte, *i.e.*, the ionic conductivity, the electrochemical stability window or the lithium transport number. In fact, the upscale of materials from a laboratory curiosity to a practical technology requires careful evaluations also in view of the availability of raw materials and their scarcity: in Table 1, we illustrate how performance metrics, such as ionic conductivity, can be compared to qualitative descriptors of the SE sustainability, that is the nature of the rarest atomic species and occurrence of CRMs in the electrolyte formulation. In the following, a brief overview of several SE options will be presented, with special highlights on limitations in their effective exploitations and eventual room of improvements for future sustainable assessments.

***Table 1.*** *Comparison of the ion conductivity at RT vs. sustainability descriptors (rarest atomic species and presence of CRM in the SEs) for a selected list of SEs.*

| Electrolyte formulation | Ion conductivity | Rarest element | CRM |
|---|---|---|---|
| β-$Al_2O_3$ | $10^{-3}$ - $10^{-4}$ S/cm [49] | Aluminium | - |
| $Na_{1+x}Zr_2Si_xP_{3-x}O_{12}$ (NASICON) | $10^{-3}$ - $10^{-2}$ S/cm [50] | Phosphorous | Phosphorous, Zirconium |
| $Na_{0.33}La_{0.55}ZrO_3$ (NLZO) | ~ $10^{-5}$ - $10^{-4}$ S/cm [51] | Lanthanum | Zirconium |
| $Na_3OCl$ | 2 × $10^{-4}$ mS/cm [52] | Chlorine | - |
| $Na_3OBr$ | 2 × $10^{-4}$ mS/cm [52] | Bromine | - |
| $Na_3OBH_4$ | ~ $10^{-6}$ - $10^{-5}$ S/cm [53] | Boron | Boron |
| $Na_3NO_3$ | ~ $10^{-6}$ - $10^{-5}$ S/cm [54] | Nitrogen | - |
| $Na_3OCN$ | ~ $10^{-6}$ - $10^{-5}$ S/cm [55] | - | - |
| $Na_3PS_4$ | 0.01 mS/cm [56] | Sulphur | Phosphorous |
| Polyethylene oxide (PEO): NaTFSI/NaFSI | 3.55×$10^{-6}$ S/cm [57] | Fluorine | |
| Polyethylene oxide (PEO): NaTFSI/NaFSI : $AlCl_3$ | ~ $10^{-5}$ - $10^{-4}$ S/cm [58] | Fluorine | |
| Polyethylene oxide (PEO): NaTFSI/NaFSI : $BaTiO_3$ | ~ $10^{-5}$ - $10^{-4}$ S/cm [58] | Fluorine | Titanium |
| Polyethylene oxide (PEO): NaTFSI/NaFSI : CuO | ~ $10^{-5}$ - $10^{-4}$ S/cm [58] | Fluorine | Copper |
| Polyethylene oxide (PEO): NaTFSI/NaFSI : $SiO_2$ | ~ $10^{-5}$ - $10^{-4}$ S/cm [59] | Fluorine | |
| Polyethylene oxide (PEO): NaTFSI/NaFSI : $ZrO_2$ | ~ $10^{-5}$ - $10^{-4}$ S/cm [58] | Fluorine | Zirconium |
| Polyethylene oxide (PEO): NaTFSI/NaFSI : $TiO_2$ | ~ $10^{-5}$ - $10^{-4}$ S/cm [60] | Fluorine | Titanium |

** Natural abundance in Earth's crust composition for the CRM listed in the table: Ti (0.6% wt.), P (0.1% wt.), F (0.06% wt.), Zr (0.022% wt.), B (0.001% wt.),Cu (7 ppm), La (5 ppm), Br (3 ppm). Natural abundance in Earth's crust composition of Na and Li for comparison: Na (2.3%), Li (0.002%).*

The pioneering studies on β-$Al_2O_{3s}$, a ceramic system encompassing suitable 2D diffusion channels for fast $Na^+$ transport, have pushed extensive research efforts on solid inorganic electrolytes (SIEs). While ion mobility in conventional liquid electrolytes relies on the stability of the solvation shell and the dynamics of charge transfer reactions at electrolyte-electrode interfaces, ionic transport mechanisms in SIEs involve the motion of single ionic species (*e.g.*, $Na^+$) through a rigid crystalline framework. As extensively demonstrated by theoretical and experimental works, $Na^+$ hopping among energetically stable sites that are separated by certain barriers determine the microscopic migration landscape, whose connectivity is essential to enable the fast ion transport. The ionic conductivity-temperature dependence ($\sigma$ *vs.* T) obeys the Arrhenius-like equation and is strictly related to intrinsic features of the crystalline structure [61]. The Arrhenius theory of ionic transport in solids paves the way to identify the most critical phenomena that allow the tuning of ionic conductivity in SEs: (i) the increase in the concentration of vacancies or the number of interstitial ion sites (higher $n_c$ parameter) and (ii) the design of facile transport pathways within the crystal framework (lower $E_a$ values) [62]. Enabling three-dimensional migration pathways as feasible way to enhance the bulk ionic conductivity has put the NASICON (NA Super Ionic CONductors) family in the spotlight [63]. The parent material, $Na_{1+x}Zr_2Si_xP_{3-x}O_{12}$ ($0 \leq x \leq 3$), where the 3D diffusion channels are created by the corner-sharing $SiO_4$/$PO_4$ tetrahedra and $ZrO_6$ octahedra, has pioneered the design of several NASICON-type conductors obtained *via* metal substitutions [48,64].

Inspired by the large bulk conductivity and the enhanced thermal stability ensured by the inorganic crystal structures, additional oxide-based ceramics electrolytes have been investigated. Despite their wide use in LIBs, garnet-type minerals are much less common as Na-ion conductors (*e.g.*, the most popular for Li batteries is $La_3Li_7Zr_2O_{12}$, namely LLZO). Conversely, the primary investigations on the $Li_{0.33}La_{0.55}TiO_3$ (LLTO) perovskite Li-ion conductor have inspired further studies on similar Na-ion containing electrolytes. The versatile and tailorable structure of $ABO_3$ perovskite-type oxides (where A and B are cations) is the origin of their diverse and unique properties. The mechanism for alkali ion diffusion in perovskite SIEs involves ions hopping *via* vacant A-sites. Not only concentration but also ordering of the A-site vacancies has a huge impact on ionic conductivity, as disordered distribution would allow ion transport on multiple dimensionality [65]. Thus, playing element substitutions can lead to very different outcomes. Starting from its Li analogue, $Na_{0.33}La_{0.55}ZrO_3$ (NLZO) has been proposed as Na-ion SIE, with the chance to implement aliovalent doping at the A site (*e.g.*, $Sr^{2+}$) as useful strategy to increase the lattice size and enhance ion mobility [66]. More recently, the so-called anti-perovskite phases have gained great attention in the field. Compared to conventional $ABX_3$ (X refers to a halogen anion) perovskites, anti-perovskites are electronically reversed, that is cation and anion sublattices are inverted, thus they are usually referred as $X_3BA$. The structure consists of 6-fold and 12-fold coordinated B and A anionic centers, respectively, surrounded by the cation in the X site. In particular, Na-based anti-perovskites share the general formula $Na_3OA$, with Na cations lying in the X site and B/A-site being occupied by oxide and bigger anions (*e.g.*, halides - Cl, Br, I - and their mixtures). Additional beneficial properties compared to perovskites include the X-rich, and thus Na-rich, content, which has a dominant influence on ionic conductivity [67]. Two possible mechanisms have been proposed for Na migration in these materials, differing for whether the $Na^+$ hopping proceeds *via* vacancy-mediated or interstitial-dumbbells pathways [68]. Even though the latter is commonly associated with lower barriers, lattice distortion seems to be strongly correlated to conductivity and is usually considered as the leading factor. As a matter of fact, ion migration is more effectively promoted when the material undergoes enlargement of the crystal space or increase of rotational motions. Replacement of cluster ions over halides at the A site (*e.g.*, $Na_3OBH_4$, $Na_3NO_3$ and $Na_3OCN$) is shown to be largely efficient to get superionic properties, thanks to the induced lattice expansion and the increased rotational disorder, which do not jeopardize the overall thermodynamic stability [68]. Notwithstanding the promising electrochemical outcomes, oxide-based SIEs suffer from several issues that still hamper their exploitation as solid-state Na-ion conductors on a large scale. Typically, the presence of grain boundaries and high defect concentrations can lead to a locally perturbed structure, which results in large resistive migration barriers hindering the transport of mobile ions across the interface. Still, the generally poor interfacial contact with the electrode may limit their use to rigid battery designs.

The need to modulate the brittleness and hardness of SIEs that can be integrated in more flexible devices and thus embrace different portions of the market has encouraged the investigation of sulfide-based Na-ion conductors. Ionic conductivity values at RT usually reach or even overpass that of conventional liquid electrolytes, and the tunable mechanical properties chiefly make them prone for flexible device integration [69]. The employment of $Na_3PS_4$ represents a milestone along the development of thiophosphates electrolytes [70], thanks to the stabilization of the high-temperature cubic phase by crystallization of its glassy state [48]. Introduction of vacancy or interstitial defects, as well as proper doping at the P site are widely adopted strategies to further increase the ionic conductivity [48]. A variety of Sb-, Sn-, and W-substituted sulfides have also been tested as promising conductors, and even anion substitution with the more polarizable $Se^{2-}$ has been outlined as a feasible enabler of fast Na-ion transport [71]. From recent scientific reports, easy and handy cold-pressing treatments of such soft materials seem to be sufficient to ensure good contact with electrode materials during cell assembly [72]. However, the poor chemical and electrochemical stability, usually associated to decomposition and release of $H_2S$ [72], calls for safe coatings or stable passivation layers able to protect the electrode interface upon cycling, especially when used in combination with metal anodes or high-voltage cathodes [73,74].

Soft mechanical properties mainly motivate the most recent research topics focusing on complex hydrides, that are low-toxic compounds gaining great attention due to their good compatibility with metal anodes [75]. In particular, hydroborates with large cluster anions, $[B_xH_x]^{2-}$ (x = 10, 12), and their C-derivatives $[CB_{x-1}H_x]^-$, show enhanced oxidative stability owing to the strong electronic delocalization among their cages [76]. The loose crystal packing of large and highly symmetric polyhedral anions leads to weak cation-anion coordination (*e.g.*, $Na^+$-$[B_xH_x]^{2-}$) in the lattice, which in turn leaves a wide number of cation vacant sites. Moreover, the electron withdrawing effect of C-substitution in carbo-hydroborates further enhances the cation coulombic repulsion and so its motion [77]. Anions with high degrees of freedom undergo rotational motions that drive some order-disorder phase transitions and thus promote the cation mobility [78]. However, most of these phase transitions take place at high temperature, limiting the ionic conductivity and thus the incorporation of complex hydrides in solid-state batteries operating at RT. Anions mixing or nanostructuring *via* mechanochemical treatments can lower the transition temperature [76], while nano-confinement of the metal hydride into a non-conducting oxide scaffold (*e.g.*, $NaBH_4/Al_2O_3$ and $NaNH_2/SiO_2$) has been shown to enhance the ionic conductivity thanks to a highly conductive layer generated at the nanocomposite interface [79].

As a general consideration, the use of ceramic electrolytes may usually cause highly resistive interfaces with the solid-state electrode. Tuning interfacial surface contacts is a major issue in the development of versatile components and represents a remarkable challenge especially for flexible battery designs [80]. This highly pursued goal can be achieved by targeting the material softness. Solid polymer electrolytes (SPEs) play a central role to this end, claiming tailored and well-suited mechanical properties and also meeting safety and reliability requirements [81]. The good interfacial stability of SPEs can be associated with their flexible and suitable morphology that ensure optimal compatibility to the electrode surface. Thin films and membranes with self-standing ability can be easily fabricated *via* light-driven free-radical polymerization process, which modifies the microscopic network of the polymer matrix endowing highly stable cross-linked polymer electrolytes (XPEs) [82]. The sizable assortment of polymer electrolytes and the lack of simple structure-properties correlations entangle a fully comprehensive physical description for ion transport in SPEs [81]. The Vogel-Tammann-Fulcher (VTF) equation can adequately describe the ionic conductivity-temperature dependence [81]. Empirical fittings of the VTF parameters as well as computational-assisted investigations on multiple SPE systems have shed light on the ionic transport mechanisms occurring within the polymer matrix. The so-called segmental motions of mobile ions among polymer chains are responsible for ion transport across the electrolyte medium. Beyond material properties (such as viscosity or glass transition temperature), the main features affecting ion mobility are related to the energy landscape and thus act on the *B* parameter of VTF model. For example, the extent of ion-polymer coordination and the underlying bond strength sensibly affect the ion mobility in salt-polymer formulations. The relatively inert poly(ethylene oxide), PEO, is certainly the most investigated SPE. Whereas dominating the scientific literature on lithium-polymer batteries, PEO-based electrolytes also appear as a viable choice for the Na counterpart [83]. The formation of stable complexes with multiple salts makes PEO a highly promising and popular host material for both Li and Na ions [84,85]. However, the ion mobility strictly relies on its amorphous domains, with dramatic drops occurring upon crystallization of the highly symmetrical ethylene oxide (EO) units coordinating the salts [86]. The rather low conductivity of PEO-based electrolytes ($< 10^{-4}$ S • $cm^{-1}$ at RT) often requires the cells to be cycled above the polymer melting point, *i.e.*, 60-70°C. Several options have been put on the table to boost the ionic conductivity of SPEs near RT. Regulation of electrolyte salts (*e.g.*, replacing NaTFSI with NaFSI) or the polymer matrix itself (*e.g.*, shifting from PEO to polycarbonates or polyurethane) conveys common strategies to adjust ionic conductivity at RT [87]. Besides, employing liquid plasticizers that are entrapped in the polymer matrix (plasticized/gel polymer electrolytes, PPEs/GPEs, depending on whether the fraction of the liquid phase is less/more than 50% wt.) would allow to mitigate the increase of glass transition temperature induced by salt addition [84,88]. As virtually non-volatile, non-flammable and stable materials and valid alternatives to conventional and less sustainable organic solvents [89], RT ionic liquids (RT-ILs) feature large (soft) anions that would easily release the small (hard) $Li^+$/$Na^+$ from the strong EO coordination, with beneficial effect on the overall ion mobility [84,88,90,91].

As far as we have reviewed, the overall balance in the electrochemical performance of a solid-state sodium battery is clearly determined by the chemical state of the SE and the underlying mechanisms involved for the Na-ion transport. The use of ceramic-polymer composite materials may combine the advantages of both components while overcoming their individual drawbacks (*i.e.*, balancing RT ionic conductivity and electrode interfacial stability, as shown in Fig. 2). Composite polymer electrolytes (CPEs) can be designed as polymer matrix encompassing an inorganic phase, but also as ceramic fillers in a polymeric medium (*e.g.*, PEO, polyvinyl pyrrolidone - PVP - and polyaniline - PANI) [92]. The employed ceramic compound can be either active or passive with respect to its capability as Na-ion conductor. Some inert fillers include $AlCl_3$, $BaTiO_3$, CuO, $SiO_2$, $ZrO_2$, $TiO_2$, while the most employed conductors are NASICON-like systems. Both ways, conductive networks can be established at the interface with the polymer, with additional charge carriers eventually provided by active fillers that directly participate to ion transport [93]. The ceramic-polymer interface is believed to play a key role in the formation of amorphous-rich PEO regions and the increase of mobile ions concentration at the interfaces [93]. The open debate in the literature concerning possible transport mechanisms in CPEs mainly ascribe the conductivity enhancement to whether the inhibition of polymer crystallization or the optimized interfacial contacts [94]. It is reasonable to consider the ion transport occurring both *via* segmental motions within the amorphous polymer regions and through an activated hopping mechanism enabled by the presence of additional vacancies on the ceramic surface. Not only ionic conductivity, but also the overall mechanical stability of CPEs is shown to benefit from such synergistic effects.

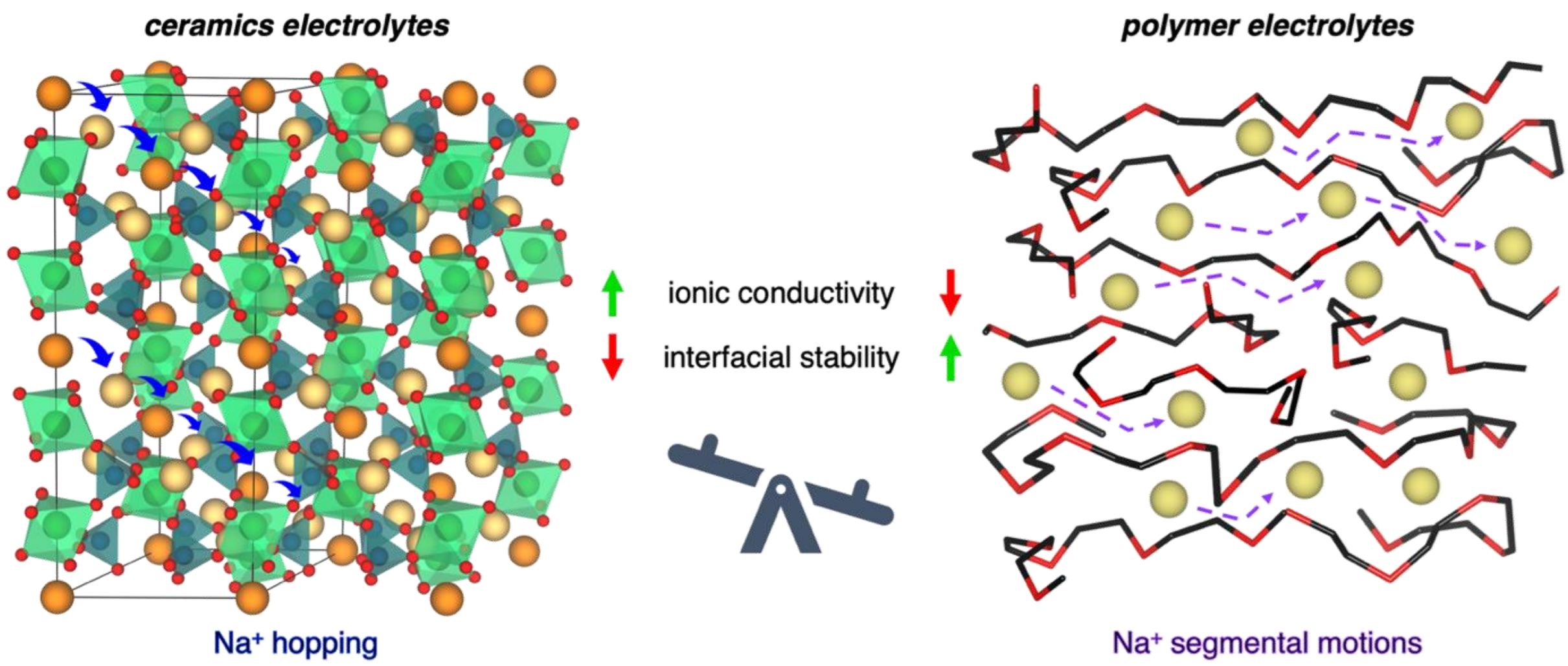

***Figure 2.*** *Pictorial representation of $Na^+$ transport mechanisms in ceramics and polymer electrolytes as the fundamental feature determining the overall balance of electrolyte performances, i.e., the ionic conductivity and the interfacial stability.*

All-in-all, the deployment of all-solid-state Na batteries is still hindered by deficiencies in SEs and poor contact with large interfacial resistances at solid electrolyte-electrode interfaces, which have highly unfavorable effects on the overall battery performances. Critical issues have been addressed by implementing advanced design strategies. Tailored substitutions or doping in SIEs, as well as blending, copolymerization, and cross-linking techniques in SPEs are highlighted as viable solutions to improve ionic conductivity and electrochemical stability. On the other hand, interfacial stability issues, usually resulting in large resistive interfaces and poor compatibility between electrode and electrolyte, can be unraveled by combining polymer and ceramics in the production of highly promising CPEs.

It is worth mentioning that acquiring the fundamental understanding of ionic conduction mechanisms in bulk materials or across the interfaces is much more demanding than assessing each single property. The design and development of new solid electrolytes with proper structures and superior properties should not disregard the microscopic behavior of the chemical systems upon battery functioning. Adopting multiscale approaches, from theoretical calculations to experimental investigations and advanced characterizations, can make the atomistic insights available, thus providing innovative design strategies able to boost such technology transition. This perspective would grant the rational development of functional materials and lead to significant breakthrough in the optimization of solid-state Na batteries in the near future.

It is important to underline that the current push towards innovative electrolyte formulations should not disregard the possible correlation/anticorrelation of electrochemical performance and environmental impact of such SEs. As we have punctually reported in Table 1, most of electrolyte formulations are strictly relying on non-innocent amounts of CRMs or otherwise rarest elements. The use of phosphorous, fluorine, zirconium, boron, in a large variety of SEs, from ceramics (NASICONs family, anti-perovskites) to polymers (especially due to F-containing salts and/or additives to produce the composite counterpart), raises concerns on the sustainability promises of ASSBs and sensibly counteracts the regulatory PFAS directive. Whenever a new high-performing electrolyte will gain any targeted ionic conductivity values, the sustainability of the overall device might be severely affected if it is done at the expense of ecofriendly, naturally abundant and cheap raw materials in the formulation.

## 4. Addressing technology validation: practical strategies towards effective solid-state interfaces

Despite the early-stage advances in solid-state Na-ion conductors, a remarkably high number of formulations have already proven to be robust and efficient in ASSB devices. A visual overview of various cell configurations is reported in Fig. 3, as available in the recent literature.

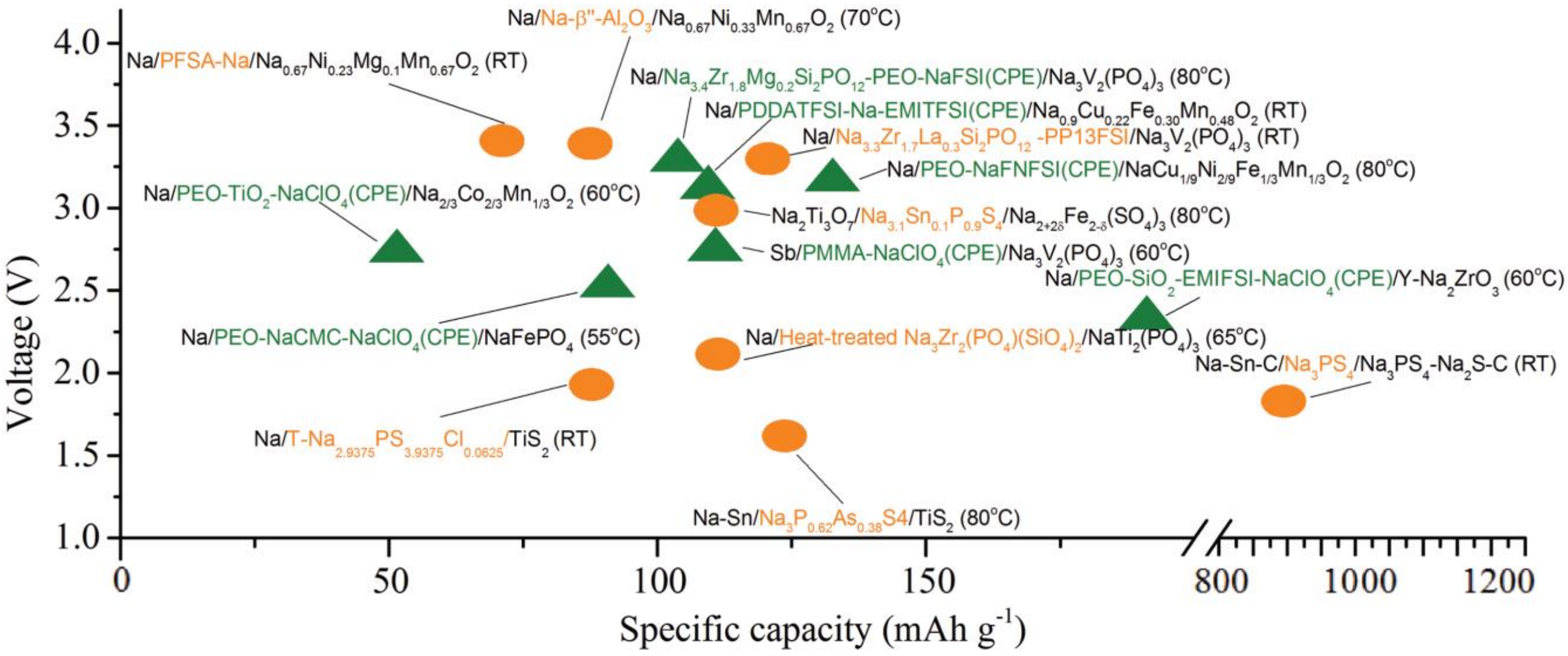

***Figure 3.*** *Voltage vs. specific capacity plot of ASSBs reported in the recent literature at specific temperature conditions. Reproduced with permission from Ref. [95] © 2025 Wiley.*

In spite of this, the industrial upscale and commercialization of ASSBs is yet negligible, mainly owing to several technological drawbacks affecting both electrodes and electrolytes, including conductivity, chemical compatibility, and storage mechanisms. The rational design of ASSBs components nowadays represents a highly desirable and fruitful strategy to develop advanced, high-energy and long lifecycle devices [96]. In addition to this, a comprehensive *in situ*, *in operando* and *post-mortem* characterization of morphological and structural features of all the components through a variety of techniques seems to be crucial for the continuous improvement of the electrochemical performances in terms of durability and safety, giving also the possibility to elucidate some mechanistic aspects [97–99]. As anticipated, not only SEs would allow getting rid of flammable and unsafe liquid electrolytes, but they can also be coupled to high-voltage cathodes thanks to the higher stability [100]. It is crucial to ensure good adhesion between electrodes and the electrolyte in order to facilitate the ion transfer and minimize the interfacial resistance. Morphological changes upon cycling are therefore more critical compared to the liquid electrolyte cases, since internal fractures can originate from the materials mismatch and lead to detrimental capacity decay. Moreover, electrolyte-electrode interface conductivity highly affects the battery life cycle and performances, consequently being considered the actual bottleneck for the development of ASSBs.

One of the possible strategies to improve the adhesion and stability of the interfaces between electrodes and electrolyte is the insertion of additional interlayers to enhance the morphological integrity of the device. In this respect, different approaches have been proposed for positive (cathode) and negative (anode) intercalation electrodes as well as the sodium metal one. The use of a ferroelectric layer between solid polymeric electrolytes and the electrodes (*i.e.*, cathode and sodium metal) reduces interfacial resistances. Basically, this layer suppresses the solid electrolyte interface (SEI) growth at the anode/electrolyte interface and enhances the ion transfer between electrolyte and cathode, improving both cell stability and capacity retention [101]. Electron-blocking interlayers based on metal oxides and salts deposited between the metallic sodium and the solid electrolyte can disclose at the same time: (i) a close contact between electrolytes and sodium metal; (ii) a fast $Na^+$ transmission; (iii) a uniform sodium deposition/dissolution over cycling, thus preventing the formation of sodium dendrites at room-temperature [102,103].

Mechanical properties and interfacial contacts can be improved by employing polymer interlayers [104]. A typical example is represented by polydopamine (PDA), which proved to have powerful adhesion to NASICON-based electrolytes and $FeS_2$-based cathodes, the latter being characterized by high theoretical capacity and low cost, but undesirably large volume change over cycling. PDA is chemically stable, easy to synthesize and flexible enough to tolerate volumetric change, enlarge the interfacial contact, and improve the structural integrity upon cycling. Its presence allows to eliminate the gaps between the electrolyte grains, while an efficient $Na^+$ transfer is obtained by exploiting the charge-transfer interactions between o-benzoquinone and catechol units [105]. Employing polymer interlayers and electrolytes also confers to the systems enhanced flexibility, which can also be achieved upon addition of even trace amounts of ionic liquids [106,107]. From a technological point of view, flexibility is an essential feature, since it allows the battery integration in a number of devices with different shapes, dimensions and scopes: for example, the use of highly flexible matrices like carbon cloth as the electrode substrate opens new perspectives in the development of smart wearable electric devices [108]. The use of suitable polymers could in principle extend the application of ASSBs to broader ranges of environmental conditions. For example, a perfluorinated sulfonic acid (PFSA)-Na membrane recently fabricated starting from PFSA-Li by a facile large-scale ionic-exchange strategy showed promising performance as SE, including high ionic conductivity in a wide temperature range, excellent thermal and mechanical stability, and a rough and porous structure which guarantees a tight contact with the electrodes [109]. Solid-state SBs assembled by coupling this polymer with a Prussian blue-based cathode not only showed remarkable cycling stability, rate capability, and coulombic efficiency at RT, but also superior cycling performance with respect to the liquid counterpart at temperatures far below 0°C [109]. The same membrane coupled with the non-toxic and cost-effective pyrophosphate polyanionic cathode, $Na_7Fe_{4.5}(P_2O_7)_4$@C, resulted in a device with outstanding performance in terms of rate capability and cycle lifespan, *i.e.*, at least 1000 cycles with a capacity decay rate of about 0.0069% per cycle [110]. For sake of completeness, one should also recall that the enhanced flexibility obtained by using polymers must balance also with their electrochemical/chemical stability and the lower conductivity compared to SEs. These drawbacks can be tackled by an increase in the minimal operation temperature. However, it is worth mentioning that using thin films can be scarcely extendable to large-scale applications because of intrinsic limitations in the mass loading. Chemical infiltration combined with *in situ* electrode materials synthesis sounds resolutory, as reported for few cathodes, like $Na_3V_2(PO_3)_4$ (NVP) on $Na_{3.4}Zr_2Si_{2.4}P_{0.6}O_{12}$ (NZSP) solid electrolyte [111,112].

In addition to these practical approaches, fundamental studies are crucial to investigate the effect of electrode/electrolyte interface morphology and crystallography in view of designing next-generation ASSBs without any additional components. Crystallographic orientation, faceting, and surface microstructure have a significant impact on the performances of solid-state batteries. Moreover, the combination of dense and hard ceramic cathodes with soft solid electrolytes offers exceptional fabrication advantages over conventional solid-state battery systems in terms of enhanced ion transport and reduced capacity fading, thanks to an improved control of chemical and electrochemical reactions at the cathode/electrolyte interface [113]. Overall, the development of optimized and effective interfaces can be achieved only by an in-depth comprehension of both storage and transfer mechanisms [114]. In general, similarly to the Li-ion case, electrochemical sodiation/desodiation processes can be attained *via* three mechanisms, *i.e.*, intercalation [115,116], conversion [117,118], and alloying [118,119]. In any case, it is crucial to keep the electrode materials expansion during the sodiation/desodiation process below 8%, since a larger volume change would cause relevant mechanical stress, which rises up to a rapid capacity fading [120].

Intercalation, in which ions are de-inserted or inserted over cycling into the electrode material, is nowadays the preferred choice for secondary batteries. In principle, the electrode forms a framework with specific sites able to host the ions, but the actual resulting electrochemical performances are largely dependent on both thermodynamic and kinetic properties of the material, which are sometimes difficult to predict and quantitatively evaluate [121–123]. Most of the intercalation systems used as cathode in SBs can be classified as layered materials and polyanion compounds. Sodium layered oxides have been studied and developed as analogues of highly exploited cathodes for LIBs. $Na_xMO_2$ materials (where M is a transition metal) are usually classified as P2 and O3 type according to (i) the prismatic or octahedral structural arrangement of the site occupied by $Na^+$, and (ii) the number of $MO_2$ layers and intercalated sodium in the repeat unit perpendicular to the layering. Sodium extraction from P2 and O3 type materials is usually coupled with a phase transition, typically to O2 and P3 respectively. Even though these materials seem rather promising as high-capacity cathodes for SBs, it is worth noticing that their hygroscopic character and sodium deficiency (especially for P2) reduce their practical use at large scale, due to the impossibility of handling them in moist air and the initial charge/discharge capacities discrepancy between the electrodes of the resulting cells [124]. To overcome the latter issue, it is possible to add a sacrificial salt able to electrochemically and irreversibly decompose by releasing $Na^+$ [125,126]. P2-$Na_xCoO_2$, with $0.6 \leq x \leq 0.74$, exhibits a certain reversibility as a cathode, but its properties largely depend on the preparative conditions. For example, when synthesized *via* sol-gel technique and having x = 0.70, it showed the typical plateau region of discharge profile up to 4 times longer and an enhanced discharge capacity in a solid-state configuration featuring a polymer electrolyte and Na metal anode, compared to the same material obtained *via* solid-state reaction or high-energy ball milling [127]. Such enhancement has been attributed to the larger surface area of the material obtained by using sol-gel methodology. However, the whole system suffered from a low value of $Na^+$ diffusion coefficient and low ionic conductivity of the electrolyte if compared to conventional liquid ones [128]. A valid alternative would be the use of high sodium content P2-type cathodes like $Na_{0.85}Li_{0.12}Ni_{0.22}Mn_{0.66}O_2$, which proved to ensure fast $Na^+$ mobility and small volume variation, together with remarkable electrochemical properties, *i.e.*, high reversible capacity, rate capability and capacity retention [129,130]. As predicted by DFT calculations and *ab initio* molecular dynamic simulations, and confirmed by experimental evidence, traditional oxide cathodes can be coupled with aliovalent substituted halide-based ionic crystal $Na_3YCl_6$, such as $Na_{3-x}Y_{1-x}Zr_xCl_6$. Conversely to their Li halides analogues, the parent $Na_3YCl_6$ compound shows negligible ionic conductivity. However, the substitution with $Zr^{4+}$ causes an increase of the volume of the unit cell, resulting in polyanion rotation and consequent formation of an interconnected network of $Na^+$ diffusion channels, which in the end determines a significant enhancement of $Na^+$ conductivity [131]. It is worth noting that the $P2_1/n$-type structure is kept up to x = 0.875, while for further substitution an additional hexagonal phase with P-3m1 space group is observed. The resulting cathode-solid electrolyte composite proved to be highly stable, even in a wide oxidative electrochemical window [23]. Sustainability represents a major issue when it comes to cathode materials containing transition metals, due to the reliance of most intercalation compounds on nickel electrochemistry [132]. The extremely low natural abundance on Earth's crust (0.0084% by weight), the sensible geopolitical factors and the intense industrial demand support the identification of Ni as CRM. The use of cobalt has raised severe concerns especially for LIBs, while it has been largely reduced in Na-based technologies [133]. Despite being more abundant (0.10% by weight), manganese has also been put in the spotlight for the crucial role in the battery production and the related wide fabrication of Mn-rich cathode materials. As the fourth most abundant element on Earth, iron undoubtedly represents a more sustainable choice [134].

If most cathode materials show overall good reversibility and minimal structural changes upon ion intercalation, the scenario for anodes is highly diversified. Different candidates have been investigated, including carbonaceous materials, metals, oxides, sulfides and their composites, aiming to substitute the conventional yet ineffective graphite electrodes for Na-ion storage [135,136]. On the other hand, amorphous hard carbons with larger interlayer spaces and different micro- and nanostructures proved to be more applicable. Typically, the sodium storage process in hard carbon-based materials is characterized by high overpotential during the de-insertion process with a sloping capacity, and a long plateau capacity during the insertion process at low-potential, which is supposed to be crucial to obtain high-energy-density SBs. Even though a huge amount of experimental evidence and systems is available, the mechanistic insights for sodium storage in carbon-based anodes are still debated, and the actual involved steps remain controversial, probably due to the variety of possible structures, which makes difficult both comparisons and systematic analysis [137,138]. Over the last 20 years, both experiments and simulations have pointed out different and conflicting mechanisms, *i.e.*, the so-called "insertion-adsorption (filling)" and the "adsorption-intercalation (insertion)" mechanism, in which the high potential sloping region and the low potential plateau region were attributed respectively prior to the $Na^+$ insertion into carbon layers and then to the $Na^+$ adsorption onto defect sites and nanopores or *vice versa* [139–141]. Considering also the presence of meso- and micropores to be filled, more recently the pure or hybrid "adsorption-intercalation-filling” mechanisms have also been proposed, in which competition between interlayer intercalation and pores filling in hard carbon is taken into account, and the overall process is highlighted to largely depend on the structure of the considered material [142,143]. A comment on the sustainability of carbon-based anode materials is required, even though it rather pertains the production and manufacturing procedures than the formulations themselves. Ecofriendly synthesis routes for hard carbons foresee the pyrolysis of waste materials, including hydrocarbons, cellulose, lignin, coconut shells or sugar, with no impact on the carbon footprint. However, the final formulations usually couple the so-prepared active material with fluorinated binders, such as the popular polyvinylidene fluoride (PVdF) [137,138]. A viable greener alternative, namely carboxymethyl cellulose (CMC), is gaining increasing attention as F-free binder. In other cases, the fabrication of highly performing hard carbons encompasses PFAS emissions, for example B- and P-doping strategies aiming to induce larger interlayer spacing and wider accommodation sites for Na ions, thus directly enhancing the anode specific capacity [137,138].

Despite having remarkable specific capacity, alloy-based materials usually show a dramatic volume expansion, which obviously decreases their appeal for the development of advanced electrodes. However, when embedded in suitable matrices, *i.e.*, carbonaceous-based ones, their volumetric expansion is somehow controlled, and this phenomenon is even more evident if they are nanostructured. In these cases, besides reducing the pulverization caused by the mechanical stress over cycling, the ion storage is enhanced due to a synergistic effect of the two components. Composite Sn-C materials seem to be promising anodes for ASSBs, since the structural integrity of the electrode is kept over cycling, when using a NASICON-type based electrolyte. Basically, after the first cycles and the stabilization of solid electrolyte interphase (SEI), highly efficient cyclability and energy efficiency are obtained at different rates because of the good coverage of the Sn nanoparticles by the amorphous carbon. Sn is first alloyed as NaSn, then as $Na_9Sn_4$, and finally as $Na_{15}Sn_4$. Experimental evidence of $Sn^{4+}$ $3d_{5/2}$, $Sn^{4+}$ $3d_{3/2}$, $Sn^{2+}$ $3d_{5/2}$ and $Sn^{2+}$ $3d_{3/2}$ have been obtained for the discharged Sn-C anode, due respectively to the final product and the intermediate partially unreacted material, respectively [144]. Tin-based alloys can be considered a borderline option in terms of sustainability. Although being relatively abundant (0.23% by weight) compared to other CRM, Sn mining is geopolitically restrained and its supply chain made quite vulnerable by the high demand in many industrial applications.

The possibility of using sodium metal is still under debate. On the one hand, exploitation, especially in large-scale production, should be avoided due to the high chemical reactivity and the relatively low melting point. On the other hand, as a soft, ductile and malleable metal, it could improve the anode wettability and interfacial contact with the electrolyte, while eventually forming weak dendrites and limiting pulverization over cycling. To conclude the environmental assessment for each class of electrode materials reviewed so far, sodium metal anode surely goes beyond the discussion. As an intrinsically sustainable material, sodium is not highlighted as CRM due to its abundance and stable supply chain. Additionally, its production *via* chloride electroplating remains quite affordable. Convenient and applicable use of sodium metal at commercial level seems to be enabled by exploiting 3D structures that are able to protect the metal and enhance the interface stability [145]. To some extent, this approach is similar to the one used for alloying materials-based anodes [146]. As an alternative, it is possible to add an interfacial interlayer between the metal and the electrolyte [147]. The presence of the interlayer limits the volume expansion and contraction of the anode over cycling, thus guaranteeing a longer life cycle to the system. Moreover, since the sodium metal wets better the interlayer than the most commonly used ceramic electrolytes, the interlayer favors a more uniform sodium flux, lowers the interface resistance, and successfully inhibits the dendrite formation [148].

Another option is the use of 2D materials, with the twofold advantage of having large specific surface area and reduced electrode thickness. Within this context, a Na-$Ti_3C_2T_x$ composite anode (where $T_x$ indicates -O, -OH, -F terminating groups) proved to be highly convenient both in terms of design and performances. Actually, having only 46% thickness of sodium metal anode, it allowed to reduce costs and increase energy density. Besides the easy fabrication, desirable mechanical properties, and chemical stability typical of $Ti_3C_2T_x$, the composite anode showed lower nucleation overpotential and enhanced stability compared to metallic sodium over repeated cycling. Coupling this composite anode in a coin cell with a PVdF-HFP-$Na_3Zr_2Si_2PO_{12}$ modified polyimide membrane and sodium metal as counter electrode has resulted in improved mechanical strength, enhanced sodium migration, increased discharge specific capacity of 89.7 mAh/g at a current density of 0.2C, and excellent capacity retention up to 500 cycles [149].

## 5. Solid-state concepts towards conversion-based cells: the future of Na-S and Na-$O_2$ batteries

Na-sulphur (Na-S) and Na-oxygen (Na-$O_2$) batteries are based on conversion positive electrodes, *i.e.*, sulphur and oxygen, respectively, instead of insertion/de-insertion positive electrode materials like conventional sodium-ion cells. Thanks to three main drivers, they are attracting great interest for grid-scale stationary energy storage applications:

(i) The abundance of sulphur and oxygen.

(ii) The feasibility to develop cost-effective and sustainable batteries.

(iii) The high theoretical specific energy density that can be achieved by combining sodium with sulphur (1274 Wh • $kg^{-1}$, with $Na_2S$ as discharge product) and oxygen (1600 Wh • $kg^{-1}$, with $Na_2O_2$ as discharge product) [150–153].

The most common Na-S and Na-$O_2$ cell configurations are based on the use of Na metal as anode, sulphur-carbon composite or porous electrode exposed to the environment (gaseous oxygen electrode) as cathode for Na-S and Na-$O_2$ cells, respectively, and aprotic organic liquid electrolyte [154]. Table 2 shows a comparison between theoretical values of cell voltages, specific capacity, and energy densities of Na-S and Na-$O_2$ cells featuring Na metal as negative electrode.

***Table 2.*** *Theoretical values of cell voltage, specific capacity and capacity density referred to the weight and density of active materials, gravimetric and volumetric energy without and including the oxygen weight. All the values refer to the discharged state. Thermodynamic data derived from HSC Chemistry for all compounds being in their standard states at 25℃* [150].

| Cell reaction | Cell voltage $E^0$ / V | Specific capacity $Q_{th}$ / mAh $g^{-1}$ | Capacity density $Q_{th}$ / mAh $cm^{-3}$ | Gravimetric energy density $E_{th}$ / Wh $kg^{-1}$ | Volumetric energy density $E_{th}$ / Wh $L^{-1}$ |
|---|---|---|---|---|---|
| $2Na + \frac{1}{2}O_2 \underset{Charge}{\overset{Discharge}{\rightleftharpoons}} Na_2O$ | 1.95 | 867 | 1968 | 2273/1687 | 3828 |
| $2Na + O_2 \underset{Charge}{\overset{Discharge}{\rightleftharpoons}} Na_2O_2$ | 2.33 | 689 | 1936 | 2717/1602 | 4493 |
| $Na + O_2 \underset{Charge}{\overset{Discharge}{\rightleftharpoons}} NaO_2$ | 2.27 | 488 | 1074 | 2643/1105 | 2341 |
| $2Na + \frac{1}{8}S_8 \underset{Charge}{\overset{Discharge}{\rightleftharpoons}} Na_2S$ | 1.85 | 687 | 1245 | 1273 | 2364/1580 |
| $2Na + \frac{1}{2}S_8 \underset{Charge}{\overset{Discharge}{\rightleftharpoons}} Na_2S_4$ (25°C) | 2.03 | 308 | 653 | 626 | 1326/997 |
| $2Na + \frac{1}{2}S_8 \underset{Charge}{\overset{Discharge}{\rightleftharpoons}} Na_2S_4$ (300°C) | 1.90 | 308 | 653 | 583 | 1124/845 |
| Na-ion (average positive electrode vs. $Na^+$/Na) | ~3.3 | ~170 | ~760 | ~560 | ~2500 |

Although the technological readiness level (TRL) of both technologies is still far from real applications (*i.e.*, TRL 1-3), many research efforts are recently focusing on the development of Na-S and Na-$O_2$ cells employing non-flammable, sodium-conducting, quasi-solid or solid-state electrolytes with the aim to improve their electrochemical performance in terms of energy density and cycle life and overcome the safety issues related to the use of liquid electrolyte. In Na-S cells, the use of SE can improve the electrode/electrolyte interface, prevent the diffusion of polysulphides (PSs), and reduce the sodium dendrite formation. In Na-$O_2$ cells, the replacement of liquid electrolyte with a solid-state one can reduce the electrolyte decomposition, thus preventing the formation of insoluble by-products that are detected on the cathode surface together with the main discharge products of oxygen reduction.
In the following sections, an overview of the state-of-the-art quasi-solid- and all-solid-state Na-S and Na-$O_2$ batteries is reported.

The first Na-S cell prototype was developed in 1960s and operated at high temperatures (300-350°C). It featured molten active electrode materials, *i.e.*, sodium metal as anode, sulphur impregnated on a graphite felt for electron exchange as cathode, and sodium-conducting β''-alumina ($NaAl_{11}O_{17}$, BASE) ceramic electrolyte that approaches the high ionic conductivity of $H_2SO_4$ aqueous electrolyte at 300°C [155]. Although the great advantages of high power, energy densities, and long cycle life, the high operating temperatures of 270-350°C also cause safety, reliability and maintenance issues that limit their widespread adoption for stationary energy storage applications. Moreover, at such high operating temperatures, molten sodium, sulphur and polysulphide compounds are highly corrosive, and a fraction of energy is required to maintain such high temperatures, thus leading to a low overall efficiency of 87% [156]. Therefore, to overcome these issues the scientific community searched for Na-S batteries operating at RT and safer operating conditions. The first RT Na-S cell was proposed in 2006, when Park *et al.* developed the first solid-state Na-S cell delivering an initial specific discharge capacity of ~ 500 mAh • $g^{-1}$. The cell, consisting of sodium metal anode, solid-composite-type sulphur electrode cathode, and a GPE based on PVdF, tetra-glyme plasticizer and sodium triflate ($NaCF_3SO_3$) as salt, featured a sodium ion conductivity of $5.1 \times 10^{-4}$ S • $cm^{-1}$ at 25°C [157].

The theoretical gravimetric energy of RT Na-S cell is 954 Wh/kg, higher than that delivered by high-temperature Na-S cell that cannot be discharged into the solid-phase sodium sulphur materials. The basic reaction mechanism at RT of a Na-S cell is based on electrochemical conversion reactions between sodium ions and sulphur, resulting in a series of long-chain PS intermediates ($Na_2S_x$, $4 \leq x \leq 8$, $Na_2S_x$, $1 \leq x < 4$) [158]. The redox reactions occurring at the positive and negative electrodes and the overall cell reaction are described in the following equations [156]:

$$2Na \leftrightarrow 2Na^{+} + 2e^{-} \qquad \text{(Positive electrode)} \qquad (1)$$

$$xS + 2e^{-} \leftrightarrow S_{X}^{2-} \qquad \text{(Negative electrode)} \qquad (2)$$

$$2Na + xS \leftrightarrow Na_2S_x \qquad E = 2.08 - 1.78\,V \qquad \text{(Overall cell)} \qquad (3)$$

Theoretically, the sodium-sulphur cell is assembled in its charged state. During the discharge process, the sodium metal is oxidized at the SEI and $Na^{+}$ are formed (Eq. 1), then they migrate through the electrolyte and reach the cathode side, where elemental sulphur ($S_8$) is reduced by accepting electrons from the external circuit (Eq. 2), and sodium PS of varying chain lengths ($Na_2S_x$) are produced (Eq. 3). The PS formation occurring *via* different steps involves different phase transitions, *i.e.*, solid-to-liquid (formation of high-order PS $Na_2S_8$, then reduced to $Na_2S_4$), liquid-to-solid (formation of short-chain PS from $Na_2S_4$), and solid-to-solid (formation of $Na_2S$ from $Na_2S_2$) [154,158–160].

Like Li-S batteries, the development of stable and high-performing Na-S cells must face technological issues and challenges to mitigate poor cycle life, low coulombic efficiency, and high capacity fading during the cell operation [158], all ascribable to: (i) poor electronic conductivity of elemental sulphur and its discharge products (such as $Na_2S$ and $Na_2S_2$); (ii) large volume change (up to 260% upon full discharge) of sulphur to solid-state short chain PSs; (iii) PS shuttle phenomenon, that takes place during the multi-step conversion reaction, associated to the solubility of PSs formed during the discharge into the electrolyte. Once long-chain PSs are migrated through the separator to the anode, they are reduced to insoluble and electronically non-conductive short-chain PSs, that diffuse back to the cathode and are oxidized again; (iv) non-homogeneous nucleation of Na metal upon charge and the following uncontrolled dendrites formation and continuous growth during cycling, which ultimately leads to breaking of the separator and short circuit of the cell. The use of solid-state electrolyte is one of the most effective strategies to overcome most of these drawbacks, including blocking the PS dissolution and shuttle, mitigating the dendrite growth and reducing the large volume expansion of sulphur, with an overall increase of cell safety and cycle lifetime (see Fig. 4).

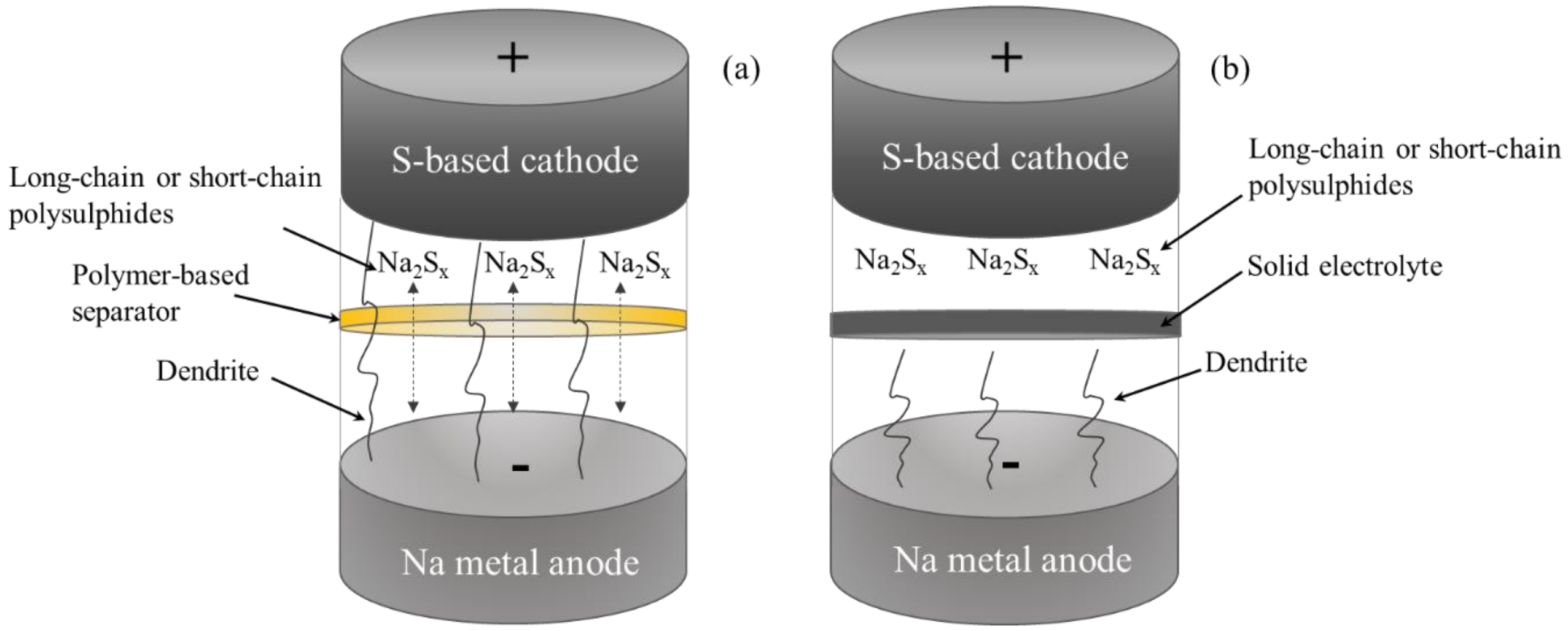

***Figure 4.*** *Scheme of a Na-S cell featuring* ***(a)*** *a conventional polymer-based separator where the PS shuttle and dendrite growth are promoted;* ***(b)*** *a solid-state electrolyte that should prevent the PS shuttle and mitigate the dendrite growth. Scheme adapted with permission from Ref.* [158], © *2025 Wiley.*

Although the research on solid-state sulphur battery concept is still in its early stage and only few studies have been reported, the promising achievements highlighted in literature encourage the battery community to increasingly conceive effective solutions [161,162]. By matching on positive and negative electrode material formulations and type of electrolyte, different solid-state Na-S battery cell designs have been developed so far, as summarized in Table 3 and discussed in the following. By adopting specific synthesis methods for the cathode material, it is feasible to develop solid-state Na-S cells approaching high specific capacity and rate capability. It has been successfully demonstrated that, to mitigate the large sulphur volume expansion over cycling, the easily scaled up thermal template method allows to fabricate metal/carbon sulphur host architecture in which Co and Ni nanoparticles are highly dispersed in carbon spheres (S@Co/C and S@Ni/C). By combining such cathodes with sodium-conductor PFSA resin as GPE, initial specific cathode capacities higher than 1000 mAh • $g^{-1}$ and capacity retention of ~ 50% after 300 cycles at 0.5C can be achieved along with stable RT cycling performance at 5C over 800 cycles [163]. High-rate performing RT solid-state Na-S cells are also enabled by combining PFSA-Na solid electrolyte with carbon-wrapped nano-cobalt anchored on graphene aerogel (S@Co/C/rGO) as cathode. The electrocatalytic effect of the nano-Co together with the conductive and flexible behaviour of graphene aerogel can provide fast chemical reaction kinetics and buffer the large volume change occurring at high

C-rates. Then, it allows solid-state Na-S cells featuring initial discharge capacity of 485 mAh • $g^{-1}$ at 0.5C and of 226 mAh • $g^{-1}$ after 1000 cycles at 5C [164]. The employment of a polymeric sulphur poly(S-penta-erythritoltetraacrylate (PETEA)-based cathode and a PETEA-tris[2-(acryloyloxy)ethyl] isocyanurate (THEICTA) as GPE with high ionic conductivity enables a RT quasi-solid-state Na-S cell featuring a specific capacity of 736 mAh • $g^{-1}$ after 100 cycles at 0.1C, and an energy density of approximately 956 Wh • $kg_S^{-1}$. The polymeric sulphur electrode effectively suppresses the PS shuttle thanks to the confinement of sulphur *via* chemical binding, besides a stable Na/GPE interface that improves the overall electrochemical performance and the cell safety [165].

RT solid-state Na-S cell featuring NASICON-like electrolyte, $Na_{3.1}Zr_{1.95}M_{0.05}Si_2PO_{12}$ (M = Mg, Ca, Sr, Ba), has been reported with alkali-earth ions doping at octahedral hexa-coordinated Zr sites through mechanochemical synthesis. Although displaying a capacity retention of 88% over 100 cycles, the cell shows quite low specific capacity (~ 170 mAh • $g^{-1}$) at 1C [44]. This can be explained considering that when moving from a GPE to a ceramic solid-state electrolyte, the ionic interface between the sodium metal and the solid electrolyte needs to be properly designed to address a good ionic path for $Na^+$ ions migration at the metal/ceramic interface. A promising approach is given by the coating of a ceramic electrolyte with a polymeric layer, as in the case of $Na_3Zr_2Si_2PO_{12}$ coated with a polymer with intrinsic nanoporosity (PIN) [166]. The PIN-coating provides an elastic buffer between Na metal anode and the solid electrolyte, thus improving the ionic interfacial properties and preventing the NASICON breaking over cycling at RT, enabling a specific capacity of about 700 mAh • $g^{-1}$ at 0.2C over 100 cycles when combined with S/carbon nanofiber (CNF) cathode.

The use of a monolithic electrolyte based on Al-doped NASICON structure ($Na_{3.5}Zr_{1.9}Al_{0.1}Si_{2.4}P_{0.6}O_{12}$) is effective for the design of RT solid-state Na metal-S cells. The porous three-layer monolithic electrolyte consists of carbon nanotubes (CNTs) and elemental S on the cathode side, and of Na metal on the anode side. The unique structure of such solid-state electrolyte enables a specific discharge capacity of about 300 mAh • $g^{-1}$ after 480 cycles at 300 mA • $g^{-1}$ [167]. Although the use of BASE electrolyte at temperatures lower than 300°C is challenging, it is feasible to develop β″-$Al_2O_3$-based solid-state Na-S operating at 80°C by designing a proper anode architecture. The employment of Na anode based on the triple $Na_xMoS_2$-carbon-BASE nanojunction interface in combination with $Mo_6S_8$/C:S@$Fe_3O_4$-NC:$PEO_{10}$-NaFSI composite cathode allows high cycling capacity of sulphur cathode of about 500 mAh • $g^{-1}$ after 50 cycles at 0.2 mA • $cm^{-2}$. The novel anode architecture confers improved elasticity for flexible deformation, intimate solid contact, and effective ionic/electronic diffusion paths, thus preventing the premature interface failure due to loss of contact over cycling [168].

Among the recently reported solid-state Na-S cells, the use of Na alloys based on alloying reactions during the sodiation/de-sodiation has been proposed as alternative to Na metal, thanks to their high theoretical specific capacities compared to those of carbon-based and Ti-based materials, while still operating at low potentials. The most promising candidates are those of the groups 14 and 15 elements, particularly Sn for the former and Sb for the latter. Despite the large volume change occurring during the full sodiation process (423% for Sn-based and 293% for Sb-based alloys), Na-alloy anode materials are effective to lower the solid interfacial resistance/instability at the anode/SEI and suppress the detrimental Na dendrite formation and growth over cycling [46,153,169]. By combining Na alloy-based anode with the glass ceramic $Na_3PS_4$ electrolyte — first synthesized as proposed by M. Jansen *et al.* — the resulting solid-state Na alloy-S cells can operate at both high and low temperatures [70]. Sulphide electrolytes have been attracting great interest thanks to favourable mechanical properties for fabricating all-solid-state batteries, high ionic conductivity, and a good processability due to a low bond energy. The lower moderate Young's modulus compared to the oxide counterpart also contributes to maintain the triple-phase solid-solid contacts among active materials, solid-electrolyte and electronic additives during volume changes over charge/discharge cycles [170,171].

Nagata *et al.* have demonstrated that an all-solid-state Na-S cell concept operating at RT can be successfully designed, featuring $Na_3PS_4$ glass ceramic electrolyte, $Na_{15}Sn_4$ as anode and a composite cathode containing S/solid electrolyte ($P_2S_5$ or $NaPS_3$ or $Na_3PS_4$)/activated carbon, albeit the cell shows a significant capacity fading between the first and the second cycle [172]. This paved the way towards the development of similar designs of all-solid-state Na alloy-S cells featuring $Na_3PS_4$ electrolyte, $Na_{15}Sn_4$ (or $Na_3Sb$)-based anode with optimized compositions and cathode composite based on sulphur, conductive carbon, and solid electrolyte. Specifically, promising cycling and rate capability performances have been reported at 60°C with Na alloy-based anode, sulphur composite cathode containing phosphorus sulphide, and $Na_3PS_4$ electrolyte over 50 [46,173] and 180 [174] cycles with specific capacities higher than 700 $mAh \cdot g^{-1}$.

Proof-of-concepts of solid-state Na alloy-S cells with $Na_3PS_4$ electrolyte have also been recently demonstrated at RT. By matching effective strategies to optimize anode and cathode formulations and interfacial engineering, it is viable to enhance the electrode/SEI and reduce the interfacial resistances over charge/discharge cycles [175–178]. However, the specific capacity and the cycling stability still need to be enhanced to make the solid-state Na alloy-S cells operating at RT comparable to those operating at higher temperatures. A sound approach to address such a challenge has been provided using $Na_3PS_4$ solid electrolyte in RT solid-state Na-S cell featuring Na metal anode and $FeS_2$/$Na_3PS_4$/acetylene black composite cathode. With this cell configuration, approximately 300 mAh • $g^{-1}$ discharge capacity after 100 cycles (at 60 mA • $g^{-1}$) and a capacity retention of 80% are achieved [179]. Significant efforts have also been dedicated also to improving ionic conductivity of these classes of sulphide solid electrolytes by carefully engineering their structure and composition. In 2012, A. Hayashi *et al.* reported for the first time the synthesis of $Na_3PS_4$ in its cubic phase instead of tetragonal through a ball milling treatment combined with an annealing of the material starting from its glassy state, reaching a conductivity of $2.0 \times 10^{-4}$ S • $cm^{-1}$ [180]. The resulting electrolyte was employed in a solid-state Na-S cell operating at RT, utilizing a Na-Sn alloy anode and $TiS_2$ cathode. The cell exhibited a specific capacity of 90 mAh • $g^{-1}_{TiS2}$, which was maintained over 10 cycles. By using a purer and crystalline $Na_2S$ precursor, the conductivity value of $Na_3PS_4$ solid-state electrolyte in its cubic phase has been doubled reaching $4.6 \times 10^{-4}$ S • $cm^{-1}$ [181]. However, the resulting Na-Sn/$NaCrO_2$ battery cell showed a lower specific capacity of 60 mAh • $g^{-1}$ over 15 cycles. The relation between the crystal phase and ionic conductivity remains complex and is still a subject of ongoing debate. High conductivity values have also been achieved also for the material in its more common tetragonal phase: S. Takeuchi *et al.*, by investigating different synthesis routes, found a tetragonal phase having larger lattice volume and an increased conductivity of $3.39 \times 10^{-3}$ S • $cm^{-1}$ [182]. The Na-S cell featuring this SE, a metallic sodium anode and a $NaCrO_2$ cathode, showed an initial specific capacity of 107.6 mAh • $g^{-1}$, which, however, rapidly decreased to 59.9 mAh • $g^{-1}$ after 10 cycles. Deeper investigations on the structure-function relationship found that $Na_3PS_4$ in its cubic phase has a tetragonal structure locally and the differences in ionic conductivity are mainly due to defects and vacancies concentration rather than different crystal structures [183]. In this context, defect engineering through aliovalent doping has proven to be a highly effective strategy for introducing vacancies and defects in a controlled manner. For instance, the partial substitution of $S^{-2}$ with halide ions leads to a less negatively charged structure that reaches a neutral overall charge with the formation of sodium vacancies [184], ultimately improving the Na diffusion inside the material. Indeed, chlorine doping has been successfully used achieving $1.14 \times 10^{-3}$ S • $cm^{-1}$ conductivity by using a $Na_{2.9375}PS_{3.9375}Cl_{0.0625}$ formulation [185]. When this electrolyte was embedded into a metallic sodium/$TiS_2$ full cell, an initial discharge capacity of 240 mAh • $g^{-1}$ was achieved. However, the capacity declined up to 80 mAh • $g^{-1}$ after 10 cycles. Bromine has also been used to substitute sulphur and increase ionic conductivity, reaching a remarkable value of 1.15× $10^{-3}$ S • $cm^{-1}$ conductivity after a hot press treatment that leads to a more compact and void-free material [186]. An initial discharge capacity of 645 mAh • $g^{-1}$ has been achieved in a solid-state full cell composed of $Na_{15}Sn_4$ anode and $Na_2S$ cathode. After 50 cycles, the cell still maintained a high reversible capacity of 420 mAh • $g^{-1}$, featuring a capacity retention of 76.4%.

Similar approaches have been developed also substituting the cationic backbone, for instance by substituting $P^{5+}$ with $Si^{4+}$ and $As^{3+}$ [187,188]: the Na-Sn/$Na_3P_{0.62}As_{0.38}S_4$/$TiS_2$ full cell operating at 80°C showed a capacity of 163 mAh • $g^{-1}$ in the first discharge cycle that stabilized at 103 mAh • $g^{-1}$ after nine cycles. It has been demonstrated also the possibility to substitute the mobile ion, introducing $Ca^{2+}$ as $Na^+$ substituent [189]. This type of doping not only promotes the formation of $Na^+$ vacancies but also affects the crystal structure of the material by stabilizing the cubic phase with localized tetragonal structure. Isovalent doping has also been used to improve the thiophosphate material performances. Indeed, a new class of oxysulfide glass has been recently developed partially substituting sulphur with oxygen as solid electrolyte for solid-state Na-S cell configuration. The bi-layer oxygen-doped $Na_3PS_{3.85}O_{0.15}$|$Na_3PS_{3.4}O_{0.6}$ solid electrolyte displays fully homogeneous glass structure and robust mechanical properties due to oxygen bridging that allows designing a solid-state Na metal-S battery with good electrochemical performance even at different current densities over 150 cycles at 60°C [190]. The tetragonal sodium superionic conductor $Na_3SbS_4$ is considered a promising solid electrolyte for solid-state Na batteries thanks to its high conductivity of 1.1 mS • $cm^{-1}$, good stability in dry air, and scalable solution processability by using methanol or water [191]. By combining $Na_3Sb$-$Na_3SbS_4$ as anode alloy, S/activated carbon MSP20-$Na_3SbS_4$ as composite cathode and $Na_3SbS_4$ as solid electrolyte, the solid-solid interfaces have been greatly improved, then the final RT solid-state Na-S cell exhibits a specific capacity of 1450 mAh • $g^{-1}$, a coulombic efficiency close to 100% with a capacity retention of 93% over 50 cycles [192]. By employing the $Na_3SbS_4$ solid electrolyte with Na metal anode and the unique design of nano-scaled S-$Na_3SbS_4$-C cathode composite, featuring 3D distributed primary and secondary ionic/electronic conduction network, the fabrication of solid-state Na-S cells with high-rate capability and high-rate cycling performance has been proved. At RT, the cell shows high discharge capacity of 1504 mAh • $g_S^{-1}$ at 50 mA • $g^{-1}$ with coulombic efficiency of 98% and a specific capacity of 662 mAh • $g^{-1}$ at 2000 mA • $g^{-1}$. When using high cathode loadings of 6.34 and 12.74 mg • $cm^{-2}$, the good electrochemical performance at 100 mA • $g^{-1}$ are still maintained, *i.e.*, 743 and 466 mAh • $g^{-1}$, respectively, thus suggesting the feasibility to develop solid-state Na-S battery cell technology in a perspective of a preliminary up-scale [179]. Furthermore, for this class of sulphide materials it is also possible to perform aliovalent doping, partially substituting antimony with tungsten. A. Hayashi *et al.* found that a $Na_{2.88}Sb_{0.88}W_{0.12}S_4$ formulation shows an outstanding ionic conductivity of $3.2 \times 10^{-2}$ S • $cm^{-1}$ after thermal treatment, due to the introduction of sodium vacancies and the stabilization of a cubic phase [193]. They also demonstrated that, by using an innovative synthetic approach based on polysulfide compounds in their liquid form, the same material could achieve an ionic conductivity of $1.25\times 10^{-2}$ S • $cm^{-1}$ [194]. The solid electrolyte was tested in a $Na_{15}Sn_4/TiS_2$ full cell containing $Na_3BS_3$ glass and $Na_{2.88}Sb_{0.88}W_{0.12}S_4$ as solid electrolyte. The cell showed a capacity of 130 mAh • $g^{-1}_{TiS2}$ after 300 cycles with a capacity retention of 76%.

The crystal structure has also been found to be important for ionic conductivity in this class of material. Indeed, H. Gamo *et al.* reported a $Na_3SbS_4$ solid electrolyte featuring crystallites in both cubic and tetragonal phase showing a RT conductivity of $3.1 \times 10^{-3}$ S • $cm^{-1}$ after ball milling of 20 h [195]. This solid electrolyte was employed in a full cell with a $Na_{15}Sn_4$ alloy anode and $TiS_2$ cathode, exhibiting a capacity of ca. 100 mAh • $g^{-1}_{TiS2}$ after 10 cycles.

***Table 3.*** *Comparison of the anode and cathode in Na-S battery cell concepts utilizing gel polymer (GPE) or solid-state electrolytes, including details on the electrolyte components and their ionic conductivity. The following acronyms are used: CB, carbon black; PFSA, Perfluorinated sulfonic resin; PETEA, poly(S-pentaerythritol tetraacrylate); THEICTA, tris[2-(acryloyl)oxy))ethyl] isocyanurate; PIN, polymer with intrinsic nanoporosity; PAN, polyacrylonitrile; BASE, β''-$Al_2O_3$; KB, carbon Ketjenblack; AB, acetylene black; VGCF, vapor-grown carbon fiber.*

| **Anode** | **Cathode** | **Electrolyte** | **Ion conductivity (S $cm^{-1}$)** |
|---|---|---|---|
| Na metal | S:Carbon:PEO (wt. 70:20:10) | PVdF-tetraglyme-$NaCF_3SO_3$ (GPE) (wt. 3:6:1) | $5.1 \times 10^{-4}$ 25°C [157] |
| Na metal | S@Co/C,S@Ni/C,S@C (wt. 6:4) CB:PVdF (wt. 80:10:10) | PFSA-Na | - [163] |
| Na metal | S@Co/C/ rGO:CB:PVdF (wt. 80:10:10) | PFSA-Na | $1.4 \times 10^{-4}$ (RT) [164] |
| Na metal | PETEA@C | PETEA-THEICTA | $3.85 \times 10^{-3}$ (25°C) [165] |
| Na metal | S/CNF | NASICON-PIN coated | $5.1 \times 10^{-3}$ (NASICON) (RT) [166] |
| Na metal | S:Super P:PVdF (wt. 80:10:10) | $Na_{3.1}Zr_{1.95}Mg_{0.05}Si_2PO_{12}$ | $3.5 \times 10^{-3}$ (RT) [44] |
| Na metal | S-CNT | $Na_{3.5}Zr_{1.9}Al_{0.1}Si_{2.4}P_{0.6}O_{12}$ | $4.43 \times 10^{-4}$ (RT) [167] |
| $Na_xMoS_2$-C-BASE | $Mo_6S_8$/C:S @$Fe_3O_4$NC: $PEO_{10}$-NaFSI (wt. 25:65:10) | BASE | $1.00 \times 10^{-3}$ (25°C), 0.25 (300°C) [168] |
| $Na_{15}Sn_4$ | S-solid electrolyte ($P_2S_5$ or $NaPS_3$-activated carbon 50:40:10) | $Na_3PS_4$ | $1.30 \times 10^{-4}$ (25°C) [172] |
| $Na_{15}Sn_4$ ($Na_3Sb$)/ KB-$Na_3PS_4$ (1:1 vol ratio) | S/KB-$Na_3PS_4$ (1:1) | $Na_3PS_4$ | $3.11 \times 10^{-4}$ (25°C), $5.50 \times 10^{-4}$ (60°C) [174] |
| Na-Sn-C | $Na_2S$/$Na_3PS_4$/ mesoporous carbon (C-MK-3) (wt. 30:40:30) | $Na_3PS_4$ | $1.43 \times 10^{-4}$ (25°C), $3.45 \times 10^{-4}$ (60°C) [46] |
| $Na_{15}Sn_4$/AB | $Na_3PS_4$/$Na_2S$-C or $Na_3PS_4$-C | $Na_3PS_4$ | $1.09 \times 10^{-4}$ (28°C), $3.40 \times 10^{-4}$ (60°C) [173] |
| $Na_{15}Sn_4$ alloy/AB | S-KB-$P_2S_5$ or S-KB-$Na_3PS_4$ | $Na_3PS_4$ | - [176] |
| $Na_{15}Sn_4$ alloy/AB | S-AB-$Na_3PS_4$ (wt. 25:25:50) | $Na_3PS_4$ | > $10^{-4}$ (RT) [177] |
| $Na_{15}Sn_4$ | S@pPAN or $Se_{0.05}S_{0.95}$ @pPAN:$Na_3PS_4$: C additive (wt. 20:60:20) | $Na_3PS_4$ | $6.9 \times 10^{-4}$ (RT) [196] |
| $Na_{15}Sn_4$/KB | $Na_2S$-NaI/ VGCF:$Na_3PS_4$ (wt. 24:10, 50:10, 70:10, 90:10, 190:10) | $Na_3PS_4$ (RT) | - [175] |
| Na metal | $FeS_2$:$Na_3PS_4$:AB (wt. 40:50:10) | $Na_3PS_4$ | $8.94 \times 10^{-5}$ (RT) [197] |
| Na-Sn | $TiS_2$ | $Na_3PS_4$ (cubic phase) | $2.0 \times 10^{-4}$ (RT) [180] |
| $Na_{15}Sn_4$ | $NaCrO_2$ | $Na_3PS_4$ (cubic phase) | $4.6 \times 10^{-4}$ (RT) [181] |
| Na metal | $NaCrO_2$ | $Na_3PS_4$ (tetragonal phase III) | $3.39 \times 10^{-3}$ (25°C) [182] |
| Na metal | $TiS_2$ | $Na_{2.9375}PS_{3.9375}Cl_{0.0625}$ | $1.14 \times 10^{-3}$ (30°C) [185] |
| $Na_{15}Sn_4$ | $Na_2S$ | $Na_{2.9}PS_{3.9}Br_{0.1}$ | $1.15 \times 10^{-3}$ (RT, after hot pressing treatment) [186] |
| Na-Sn alloy | $TiS_2$ | $Na_3P_{0.62}As_{0.38}S_4$ | $1.46 \times 10^{-3}$ (25°C) [188] |
| Na metal | S-KB-$Na_3PS_{3.85}O_{0.15}$ (wt. 2:2:6) | $Na_3PS_{3.85}O_{0.15}$ \| $Na_3PS_{3.4}O_{0.6}$ | $2.7 \times 10^{-4}$ (60°C) ($Na_3PS_{3.85}O_{0.15}$) [190] |
| $Na_3Sb$-$Na_3SbS_4$ (wt. 42:58) | S/activated carbon MSP20-$Na_3SbS_4$ (wt. 44:56) | $Na_3SbS_4$ | $5.1 \times 10^{-4}$ (25°C) [192] |
| Na metal | S-$Na_3SbS_4$-SuperP (wt. 1.2:2.0:0.4) | $Na_3SbS_4$ | $1.14 \times 10^{-3}$ (RT) [179] |
| $Na_{15}Sn_4$–KB/$Na_3BS_3$ | $TiS_2$ | $Na_{2.88}Sb_{0.88}W_{0.12}S_4$ | $1.25 \times 10^{-2}$ (25°C) [194] |
| $Na_{15}Sn_4$ | $TiS_2$ | $Na_3SbS_4$ in different crystal phases | $3.1 \times 10^{-3}$ (RT) [195] |

Turning to the Na-$O_2$ cell chemistry, it is composed of a sodium metal as anode, a porous carbon-based material exposed to air or pure oxygen as cathode and aprotic organic liquid electrolyte [150,154]. The first Na-$O_2$ cell concept was reported in 2011 by Peled *et al.*, that demonstrated the feasibility of developing a liquid-sodium-oxygen cell with polymer electrolyte and molten sodium electrode operating above 100°C [198]. During the discharge process, Na is oxidized to form sodium ions that move through the electrolyte to the cathode side where, after being combined with electrons from the external circuit, reduce the absorbed oxygen (oxygen reduction reaction, ORR). The ORR results in the formation of different sodium oxides, *i.e.*, sodium oxide ($Na_2O$), sodium superoxide ($NaO_2$), and sodium peroxide ($Na_2O_2$), while gaseous oxygen is released during charging (oxygen evolution reaction, OER). Since organic electrolytes react with $NaO_2$ or $Na_2O_2$ discharge products, the decomposition of the electrolyte takes place, thus resulting in the formation of sodium carbonates or other products on the carbon material surface. Such unwanted products electrically isolate $O_2$ with the carbon materials, leading to the fading of electrochemical performance.

Among the various approaches proposed in literature to develop stable Na-$O_2$/air batteries, the concept of quasi- or solid-state batteries is considered one of the most promising alternatives for achieving safer Na-$O_2$/air systems [199]. In this review, we focus specifically on the solid-state configurations. Solid-state Na-$O_2$ cells are still at a much earlier stage of development compared to their solid-state Na-S counterparts and remain far from practical application. In fact, research efforts are mainly focused on addressing the fundamental understanding of reaction mechanisms and the evolution of the discharge products occurring over charge/discharge cycles. This is achieved by combining special cell designs and optimized oxygen cathode formulations with advanced characterization techniques, including *in situ*, *in operando*, and *post-mortem* analysis. To shade light on the (electro)chemical and conversion reactions as well as on the discharge products, all-solid-state Na-$O_2$ cell configurations were proposed with the aim to carry out *in situ* time-resolved environmental transmission electron microscopy (ETEM) experiments and *in situ* ambient-pressure X-ray photoelectron spectroscopy (APXPS). In the former configuration, a solid-state Na-$O_2$ nanobattery is assembled within an aberration-corrected ETEM under oxygen environment, utilizing the native $Na_2O$ layer formed on the surface of the Na metal anode as the solid electrolyte [200]. In the latter, a simple and ideal model of all-solid-state Na-$O_2$ cell adopted Na metal as anode, nanoporous gold membrane as cathode and Na-β″-$Al_2O_3$ as solid electrolyte [201].

To the best of our knowledge, only few papers have showed the electrochemical performance of solid-state $Na-O_2$ cell designs. Outstanding cycling performance is demonstrated with all-solid-state $Na-O_2$ battery featuring Na metal film anode, carbon nanotube and Ru/CNT catalyst as cathode, a succinonitrile (SN)-$NaClO_4$ interlayer, and a NASICON-based solid electrolyte. The SN-$NaClO_4$ interlayer enhances the ionic conductivity and interfacial charge transfer kinetics between the three-contact interface, *i.e.*, cathode, Ru/CNT catalyst and solid electrolyte. This unique design allows remarkable electrochemical performance over 100 cycles at the current density of 100 mA • $g^{-1}$ and fixed capacity of 500 mAh • $g^{-1}$ [202]. A RT solid-state $Na-O_2$ cell has been developed by combining a NASICON-type solid electrolyte, $Na_{3.2}Hf_2Si_{2.2}P_{0.8}O_{11.85}F_{0.3}$ (NHSP-$F_{0.3}$), with NHSP-$F_{0.3}$@CNTs cathode and sodium metal anode. The cell exhibits stable cycling performance over 20 cycles and current density of 100 mA $g^{-1}$. By carrying out the electrochemical tests both in dry and wet oxygen atmosphere, it was demonstrated that the humidity had a positive impact on cycling performance. Under wet oxygen atmosphere, the formation of NaOH as discharge product is promoted. Because of the different solubility in water of NaOH and the sodium discharge products ($NaO_2$ or $Na_2O_2$) formed in dry atmosphere, the charging overpotential decreases and the reversibility of the cell notably improves, *i.e.*, 0.4 V when the cell is tested in wet oxygen atmosphere against of 0.8 V when tested in dry oxygen atmosphere [203]. The critical effect of humidity on the electrochemical performance of all-solid-state $Na-O_2$ cells has also been demonstrated in Ref. [204]. Under ~ 7% relative humidity (RH) at 80°C the all-solid-state $Na-O_2/H_2O$ cell featuring metallic sodium as anode, silver-polymer composite (SPC) as cathode and stable Na-β″-$Al_2O_3$ ceramic electrolyte is tested over repeated 100 charge/discharge cycles at 20 mA $g^{-1}$ with discharge capacity limited to 100 mAh • $g^{-1}$. The combination of SPC at the oxygen cathode and humid oxygen environment enabled good cycling performance with low overpotential of 75 mV and a round-trip efficiency of 97.1% (98.1% at the last cycle). This performance was attributed to the faster reaction kinetics of the charge process facilitated by silver and the formation of NaOH as the sole discharge product through a direct four-electron ORR, as evinced by *in situ* Raman, X-ray diffraction, and differential electrochemical mass spectrometry characterizations, and validated by DFT analysis. Although it fits to molten-salt $Na-O_2$ cell configuration, the recently investigated nitrate-mediated based $Na-O_2$ cell comprises the β-$Al_2O_3$ as membrane separator. The cell based on liquid Na negative electrode, Ni-based oxygen positive electrode, and $NaNO_3/KNO_3/CsNO_3$ eutectic salt as electrolyte is tested at 170°C under 400 cycles at 0.5 mAh • $cm^{-2}$ and 5 mA • $cm^{-2}$, displaying 33 mWh • cm-

$^{2}$ and 19 mW • $cm^{-2}$ of energy and power densities, respectively, where the dominant discharge product, $Na_2O_2$, is obtained by the nitrate-mediated ORR [151].

The use of NASICON as a solid electrolyte for Na-air batteries was recently proposed by Park *et al.* in a system operating at approximately 3.4 V. The battery relies on the reversible reactions of $Na_2CO_3 \cdot xH_2O$ (x = 0 or 1) during cycling in ambient air under varying RH conditions. They demonstrated the highest operating voltage ever reported for metal-air batteries that utilize metal oxides or carbonate/hydroxide reactions. This achievement is attributed to the *in situ* formation of a catholyte, resulting from the reaction between moisture in the air and discharge products such as NaOH. The catholyte functions both as the electrolyte and the active material, enabling reversible carbonate reactions across a large active surface area. As a result, the Na-air battery exhibits an energy efficiency exceeding 86% at a current density of 0.1 mA • $cm^{-2}$ over 100 cycles [205]. Very recently, the OXBLOLYTE project at CIC energiGUNE has been developing a solid membrane designed to mitigate oxygen crossover and enhance the stability of solid-state Na-air batteries [206].

While still at an infant stage of development, the promising achievements realized to date on both solid-state Na-S and Na-$O_2$ cell concepts should feed further fundamental studies and technology development on such battery cell designs to foster their challenging advancement towards higher TRL.

## 6. Conclusions and forward

Overall, the exploration of sodium-based solid-state batteries is advancing rapidly, driven by the pressing demand for efficient, safe, and sustainable energy storage solutions. As the market shifts towards renewable energy systems, sodium has emerged as a favourable alternative to lithium due to its abundance and cost-effectiveness. This manuscript presents a comprehensive examination of recent developments in Na-based solid-state batteries, elucidating the key materials that underpin their functionality, including innovative solid electrolytes, electrode compositions, and the mechanisms governing ion transport. The challenge of integrating these components while maintaining high energy densities and cycle stability remains an area of active research.

In summary, the advancement of solid-state sodium-based batteries represents a promising trajectory in energy storage technology. Despite facing challenges such as dendrite formation and electrolyte compatibility, continued innovations in material design and engineering are paving the way for breakthroughs in performance. The unique properties of various solid electrolytes, particularly sodium superionic conductors and polymer-based systems, demonstrate significant potential for enhancing ionic conductivity and cycle life. Furthermore, substantial improvements in electrode materials, particularly strategies that foster robust interfacial contact with solid electrolytes, exhibit a pathway to mitigating inefficiencies and promoting effective charge-discharge cycles. As the manuscript details, ongoing research into sodium electrochemical mechanisms, particularly within composite systems, will provide critical insights that can guide the development of practical sodium-based batteries. Notably, leveraging novel structural designs and optimizing existing materials is integral to overcoming current limitations. By enhancing the stability of solid-solid interfaces and maximizing the ionic mobility within solid electrolytes, there is a strong potential to achieve high-energy-density systems suitable for commercial applications. Ultimately, the insights gathered from this investigation contribute significantly to the growing discourse on sustainable energy storage, setting the stage for the eventual realization of sodium-based batteries as a competitive alternative to existing technologies in the renewable energy landscape.

## Acknowledgements

A.M., F.A.S., M.P. and S.B. would like to thank the Ministry of Environment and Energy Security for the funding in the framework of "*Ricerca di Sistema Elettrico*" – PTR 2025-27.

A.M. would like to thank the PNRR – CN1 – Centro Nazionale di Ricerca in High Performance Computing, Big Data e Quantum Computing (ICSC) – Innovation Grant project ELIO.

L.S. and F.D.G. would like to thank financial support from PNRR MUR project ECS_00000033_ECOSISTER and Prin 2022 "DiGreen: A digital and chemical approach for green recycling of Li-based batteries" (no. 2022W37L2L) — MUR (*Ministero Italiano dell'Università e della Ricerca*).